# A Study On Distributed
# Model Predictive Consensus

Tamás Keviczky* and Karl Henrik Johansson

### Abstract

We investigate convergence properties of a proposed distributed model predictive control (DMPC) scheme, where agents negotiate to compute an optimal consensus point using an incremental subgradient method based on primal decomposition as described in [1], [2]. The objective of the distributed control strategy is to agree upon and achieve an optimal common output value for a group of agents in the presence of constraints on the agent dynamics using local predictive controllers. Stability analysis using a receding horizon implementation of the distributed optimal consensus scheme is performed. Conditions are given under which convergence can be obtained even if the negotiations do not reach full consensus.

## I. Introduction

Engineered systems are becoming increasingly complex and larger in size, which presents a need for the distribution of decision-making processes that interact with or are part of these large-scale technologies and applications. An important problem that arises among such distributed decision-making systems (often called agents), is related to consensus-seeking and rendezvous, which has received a high level of interest in the recent literature [3]. The consensus-seeking and rendezvous problem consists of designing distributed control strategies such that the state or output of a group of agents asymptotically converge to a common value, a consensus point, which is agreed upon either a priori or on-the-fly using some negotiation scheme. In this paper, we assume that a consensus point is not fixed in advance, but is rather determined by an

T. Keviczky is with the Delft Center for Systems and Control, Delft University of Technology, 2628 CD, Delft, The Netherlands, `t.keviczky@tudelft.nl`

K.H. Johansson is with the ACCESS Linneaus Center, School of Electrical Engineering, Royal Institute of Technology (KTH), 100 44 Stockholm, Sweden, `kallej@ee.kth.se`

optimal control problem. We focus on the combination of model predictive controllers and subgradient-based negotiation of optimal consensus (along the lines of the work in [2]), and investigate conditions for asymptotic convergence of such distributed control schemes. We propose an algorithm for distributed model predictive consensus, which guarantees convergence under reasonable assumptions given a sufficient number of subgradient iterations can be performed without interruption.

We will model agents as constrained linear dynamical systems and build on the decentralized negotiation algorithm described in [2] to compute exactly or at least approach the optimal consensus point. This negotiation algorithm relies on primal decomposition of the optimal consensus and control problem and makes use of distributed implementation of an incremental subgradient method. Each agent performs individual planning of its trajectory and negotiates with neighbors to find an optimal or near optimal consensus point, before applying a control signal.

The paper is structured as follows. Section II introduces the optimal consensus problem and some basic notation and assumptions. The decentralized negotiation scheme of [2] is summarized in Section III along with a decentralized receding horizon implementation of the optimal consensus problem. Stability of the proposed decentralized negotiation and control scheme is studied in Section IV for both converged and interrupted negotiations. Finally, Section V presents a numerical simulation example, which illustrates the approach for an aerial refueling scenario, and Section VI the conclusions.

## II. Problem Formulation

Consider $N > 1$ dynamic agents whose dynamics are described by the following discrete-time state equations

$$
\begin{aligned}
x_{t+1}^i &= A^i x_t^i + B^i u_t^i, \\
y_t^i &= C^i x_t^i,
\end{aligned}
\tag{1}
$$

for $i = 1, \ldots, N$, where $A^i \in \mathbb{R}^{n^i \times n^i}$, $B^i \in \mathbb{R}^{n^i \times m^i}$ and $C^i \in \mathbb{R}^{p \times n^i}$. We assume that the states and inputs of each agent are constrained to lie in polyhedral sets

$$
x_t^i \in \mathcal{X}^i, \quad u_t^i \in \mathcal{U}^i, \quad t \geq 0.
\tag{2}
$$

*Definition 1:* [2] The dynamic agents described by (1) reach *consensus* at time $T$ if

$$y_{T+k}^i = \theta, \quad \forall k \geq 0, \quad i = 1, \ldots, N,$$

$$u_{T+k}^i = u_T^i, \quad \forall k \geq 0, \quad i = 1, \ldots, N,$$

(3)

where $\theta$ lies in a compact and convex set $\Theta \subset \mathbb{R}^p$.

In this paper, the consensus point $\theta$ is a vector that specifies, for example, the position and velocity the agents shall converge to.

Our objective is to find a consensus point $\theta \in \Theta \subset \mathbb{R}^p$ and a sequence of inputs $u_0^i, \ldots, u_{T-1}^i$, with $i = 1, \ldots, N$ and $u_t^i \in \mathcal{U}^i$ for all $t = 1, \ldots, T-1$, such that all agent outputs are equal at time $T$:

$$y_T^i = \theta, \quad i = 1, \ldots, N.$$

(4)

We will also require each agent to be at an equilibrium at time $T$ and denote the state and control equilibrium pairs of the $i$-th agent corresponding to a $\theta$ value with $(x_e^i(\theta), u_e^i(\theta))$. The set of equilibria for each agent $i = 1, \ldots, N$ thus will be a function of $\theta$ on the domain $\Theta$:

$$\mathcal{E}^i(\theta) = \left( x_e^i(\theta), u_e^i(\theta) \right)$$

$$= \left\{ x \in \mathbb{R}^{n^i}, u \in \mathbb{R}^{m^i} \mid x = A^i x + B^i u, C^i x = \theta \right\}.$$

(5)

We assume that the following cost function is associated with the $i$-th system:

$$V^i \left( x_k^i, u_k^i, \theta \right) = \left( x_k^i - x_e^i(\theta) \right)^\mathsf{T} Q^i \left( x_k^i - x_e^i(\theta) \right)$$

$$+ \left( u_k^i - u_e^i(\theta) \right)^\mathsf{T} R^i \left( u_k^i - u_e^i(\theta) \right),$$

(6)

where $Q^i \in \mathbb{R}^{n^i \times n^i}$ and $R^i \in \mathbb{R}^{m^i \times m^i}$ are positive definite symmetric matrices (i.e., we penalize deviations from the equilibrium states corresponding to the consensus point and the use of control effort).

*Assumption 1:* Each agent dynamics $(A^i, B^i)$ is controllable and systems $(A^i, (Q^i)^{\frac{1}{2}})$ are observable.

We then formulate the following finite-time optimal control problem at time $t$ based on [2]:

*Problem 1:* Let $T > 0$ be fixed. Determine control vectors $u_{k,t}^i$, $k = 0, \ldots, T-1$, for all

$i = 1, \ldots, N$ and the consensus point $\theta_t$, which solve the following optimization problem:

$$\min_{U_t, \theta_t} \quad \sum_{i=1}^{N} \sum_{k=0}^{T-1} V^i \left( x_{k,t}^i, u_{k,t}^i, \theta_t \right)$$

$$\text{subj. to} \quad x_{k+1,t}^i = A^i x_{k,t}^i + B^i u_{k,t}^i, \tag{7a}$$

$$y_{k,t}^i = C^i x_{k,t}^i,$$

$$x_{k,t}^i \in \mathcal{X}^i, \quad k = 1, \ldots, T, \tag{7b}$$

$$u_{k,t}^i \in \mathcal{U}^i, \quad k = 0, \ldots, T-1, \tag{7c}$$

$$y_{T,t}^i = \theta_t, \tag{7d}$$

$$x_{T,t}^i = x_e^i(\theta_t), \tag{7e}$$

$$x_{0,t}^i = x_t^i, \tag{7f}$$

$$i = 1, \ldots, N,$$

$$\theta_t \in \Theta, \tag{7g}$$

where $U_t \triangleq [u_{0,t}, \ldots, u_{T-1,t}] \in \mathbb{R}^{T \sum_i m^i}$ with $u_{k,t} \triangleq [u_{k,t}^1, \ldots, u_{k,t}^N]$, denotes part of the optimization vector containing control inputs, $x_{k,t}^i$ denotes the state vector of the $i$-th agent predicted at time $t + k$ obtained by starting from the state $x_t^i$ and applying to system (1) the input sequence $u_{0,t}^i, \ldots, u_{k-1,t}^i$. The full optimization vector consists of the vector $U_t$ defined above and the consensus variable $\theta_t$. The subscript $t$ will be significant later in Section III, when this problem will be solved repeatedly in a receding horizon fashion.

By implementing the solution to Problem 1, agents reach consensus at time $T$ in the sense of Definition 1. We will make the following assumptions on the feasibility of reaching the consensus point by all agents:

*Assumption 2:* The rendezvous time horizon $T$ is large enough so that all $\theta_t$ in the set $\Theta$ are feasible, i.e., reachable consensus equilibrium points for all agents.

*Assumption 3:* For all $\theta_t \in \Theta$ and $i = 1, \ldots, N$, there exists a sequence $u_0^i, \ldots, u_{T-1}^i$ in the relative interior of $\mathcal{U}^i$ such that $y_T^i = \theta$.

This means that it should be possible to reach $\theta_t$ without saturating the control signal (not necessarily in an optimal way).

The solution of Problem 1 was distributed among the agents in [2] by using primal decomposition in combination with an incremental subgradient method [4]. First, a multiparametric solution of the individual optimization problems was defined as

$$q^i(x_t^i, \theta_t) = \min_{U_t} \quad \sum_{k=0}^{T-1} V^i\left(x_{k,t}^i, u_{k,t}^i, \theta_t\right) \tag{8}$$

$$\text{subj. to} \quad (7a) - (7g), \quad k = 1, \ldots, T-1.$$

The optimal consensus problem in (7) can then be written as

$$q^*(x_t) = \min_{\theta_t} \quad \sum_{i=1}^{N} q^i(x_t^i, \theta_t) \tag{9}$$

$$\text{subj. to} \quad \theta_t \in \Theta.$$

The set of optimal consensus points is defined as

$$\Theta_t^* = \left\{\theta_t \in \Theta \;\middle|\; \sum_{i=1}^{N} q^i(x_t^i, \theta_t) = q^*(x_t)\right\}. \tag{10}$$

It can be established that the cost function $q^i(\cdot)$ defined in (8) is a convex function and a subgradient $g^i$ for $q^i(\cdot)$ at $\theta_t$ is given by the Lagrange multipliers corresponding to the terminal point constraint.

A principal method for solving problem (8) is the subgradient method

$$\theta_t(k+1) = \mathcal{P}_\Theta\left[\theta_t(k) - \alpha(k)\sum_{i=1}^{N} g^i(k)\right] \tag{11}$$

where $g^i(k)$ is a subgradient of $q^i$ at $\theta_t(k)$, $\alpha(k)$ is a positive stepsize, and $\mathcal{P}_\Theta$ denotes projection on the set $\Theta \subset \mathbb{R}^p$. In the following, we will consider the incremental subgradient method proposed in [5]. It is similar to the standard subgradient method (11), the main difference being that at each iteration $k$, $\theta_t$ is changed incrementally, through a sequence of $N$ steps. Each step is a subgradient iteration for a single component function $q^i$, and there is one step per component function. Thus, an iteration can be viewed as a cycle of $N$ subiterations. If $\theta_t(k)$ is the vector obtained after $k$ cycles, the vector $\theta_t(k+1)$ obtained after one more cycle is

$$\theta_t(k+1) = \vartheta_t^N(k), \tag{12}$$

where $\vartheta_t^N(k)$ is obtained after the $N$ steps

$$\vartheta_t^i(k) = \mathcal{P}_\Theta\left[\vartheta_t^{i-1}(k) - \alpha(k)g^i(k)\right],$$
$$g^i(k) \in \partial q^i(x_t^i, \vartheta_t^{i-1}(k)), \quad i = 1, \ldots, N, \tag{13}$$

starting with

$$\vartheta_t^0(k) = \theta_t(k), \tag{14}$$

where $\partial q^i(x_t^i, \vartheta_t^{i-1}(k))$ denotes the subdifferential (set of all subgradients) of $q^i$ at the point $\vartheta_t^{i-1}(k)$. The updates described by (13) are referred to as the subiterations of the $k$-th cycle.

We will make the following assumptions, which will allow us to formulate well-posed problems and characterize the number of subgradient iterations needed for convergence to a certain tolerance.

*Assumption 4 (Existence of an Optimal Solution):* The optimal solution set $\Theta_t^*$ is nonempty.

*Assumption 5 (Subgradient Boundedness):* There exists a scalar $\beta$ such that

$$\|g^i\| \leq \beta, \tag{15}$$

$$\forall g^i \in \partial q^i \left(x_t^i, \theta_t(k)\right) \cup \partial q^i \left(x_t^i, \vartheta_t^{i-1}(k)\right),$$

$$i = 1, \ldots, N, \quad k \geq 0,$$

where $N$ is the number of subiterations in each cycle.

Since we assume that the set $\Theta$ is compact, Assumptions 4 and 5 are automatically satisfied.

*Definition 2:* We will denote the Euclidean distance from a point $z$ to the set $\Theta_t^*$ by $\mathbf{dist}(z, \Theta_t^*)$.

*Definition 3:* A function $\gamma(\cdot)$, defined on nonnegative reals, is a $K$ function if it is continuous, strictly increasing with $\gamma(0) = 0$.

In the next section, we briefly describe the agreement mechanism of [2] and propose a closed-loop feedback control policy, which can be used in a receding horizon fashion, interleaved with subgradient-based negotiation of optimal consensus point updates.

## III. Decentralized Negotiation and Receding Horizon Implementation Scheme

The optimal consensus point $\theta_t^*$ can be computed in a distributed way using the incremental subgradient method described in (12)-(14). Reference [2] describes an algorithm, where an estimate of the optimal consensus point is passed around between agents. Upon receiving an estimate from its neighbor, an agent solves the optimization problem (8) to evaluate its cost of reaching the suggested consensus point and to compute an associated subgradient (Lagrange multiplier of terminal point constraint). The agent then performs a subiteration by updating the consensus estimate according to (13) and passing the estimate to the next agent. Each agent only computes a subgradient with respect to its own part of the objective function and not the global objective function. The convergence of the incremental subgradient algorithm is guaranteed if the agents can be organized into a cycle graph (for more details see [2]).

*Remark 1:* Besides some technical assumptions given in [1], the primal decomposition scheme and convergence to the optimal solution of (7) using sequential local subgradient iterations is possible due to decoupled and independently constrained agent dynamics. Furthermore, the overall objective function is decomposable into a sum of terms that share only a single coupling variable, $\theta_t$. Thus fixing a $\theta_t$ value in the cost and constraints separates the optimal control problem into local ones.

The control solution $U_t^*$ corresponding to a negotiated optimal consensus point $\theta_t^*$ provides an open-loop control strategy for finite-time optimal consensus. However, this solution is sensitive to model mismatch and disturbances, which suggests considering a receding horizon implementation and repeated solution of the finite-time optimal consensus problem due to its feedback nature. Our goal in such an approach is to guarantee constraint fulfillment and asymptotic convergence to a consensus point by repeatedly solving optimal consensus problems and implementing the first sample of the control solution.

More formally, let $U_t^* = [u_{0,t}^*, \ldots, u_{T-1,t}^*]$ and $\theta_t^*$ be an optimal solution of (7) at time $t$. Then, the first sample of $U_t^*$ is applied to the collection of agents:

$$u_t = u_{0,t}^*. \tag{16}$$

The optimization (7) is repeated at time $t + 1$, based on the new state $x_{t+1}$.

*Remark 2:* Stability of such a combination of DMPC and incremental subgradient methods is not a trivial question, especially since the terminal constraint value in the receding horizon scheme based on (7) is an optimization variable as well. The main point of the following investigation is to rule out a scenario where repeatedly solving and implementing the first step of a finite-time optimal control solution with changing terminal constraint value eventually results in divergence or lack of stability. Compared to the work in [1], this question arises because we are no longer considering only the open-loop implementation of a control sequence that terminates with the value $u_e(\theta_t^*)$ at time $T$, but one that is updated every time step (along with $\theta_t^*$), based on new measurements in a receding horizon fashion.

## IV. STABILITY ANALYSIS

In this section we will be primarily interested in establishing conditions for asymptotic convergence of the combined DMPC and consensus algorithm to the set of equilibria defined as

$$\mathcal{E} = \left( \mathcal{E}^i(\theta), \ldots, \mathcal{E}^N(\theta), \theta \right), \quad \theta \in \Theta. \tag{17}$$

### A. Fully Converged Negotiations

For now, we will assume that in each implementation cycle (i.e., at sampling time $t$), the distributed negotiations on the optimal consensus value $\theta_t^*$ have converged before the implementation of the corresponding control actions. In other words, the optimal solution of problem (7) is attained by every agent in each time step by means of the distributed consensus algorithm of [1]. This allows us to consider the overall system as a whole for stability analysis, using the following aggregate dynamics

$$x_{t+1} = Ax_t + Bu_t,$$
$$y_t = Cx_t, \tag{18}$$

where $A = \mathbf{diag}(A^i) \in \mathbb{R}^{\sum_i n^i \times \sum_i n^i}$, $B = \mathbf{diag}(B^i) \in \mathbb{R}^{\sum_i n^i \times \sum_i m^i}$ and $C = \mathbf{diag}(C^i) \in \mathbb{R}^{pN \times \sum_i n^i}$. The states and inputs of the overall system are constrained by

$$x_t \in \mathcal{X} = \prod_i \mathcal{X}^i, \quad u_t \in \mathcal{U} = \prod_i \mathcal{U}^i, \quad t \geq 0, \tag{19}$$

where the symbol $\prod$ denotes the standard Cartesian product of sets. Note that according to (4), consensus for the aggregate system dynamics means $y_T = Cx_T = 1_N \otimes \theta_t^*$.

Stability analysis in this case pertains to the study of the receding horizon control scheme given in (7) and (16) with a terminal point constraint to one of its optimization variables $\theta_t$. This will be performed next.

The set of states at time $k$ feasible for Problem 1 is given by

$$\mathcal{X}_k = \{ x \mid \exists u \in \mathcal{U} \text{ such that } Ax + Bu \in \mathcal{X}_{k+1} \} \cap \mathcal{X},$$

with

$$\mathcal{X}_{T-1} = \{ x \mid \exists u \in \mathcal{U} \text{ and } \theta \in \Theta \text{ such that}$$

$$x = Ax + Bu \text{ and } C(Ax + Bu) = 1_N \otimes \theta \} \cap \mathcal{X}.$$

Denote with

$$c(x_t) = u^*_{0,t}, \tag{21}$$

the control law obtained by applying the receding horizon control policy in (7) and (16) with cost function (6) for each agent, when the current state is $x_t = [x^1_t, \ldots, x^N_t]$. Consider the aggregate dynamical model (18) and denote with

$$x_{t+1} = Ax_t + Bc(x_t), \tag{22}$$

the closed-loop dynamics of the entire system. In the following theorem, we state sufficient conditions for the asymptotic convergence of the closed-loop system to the set of equilibria $\mathcal{E}$.

*Theorem 1:* Assume that

(A0)  $Q^i \succ 0, R^i \succ 0$ for all $i = 1, \ldots, N$.

(A1)  For all $\theta_t \in \Theta$ there exists a unique equilibrium $x^i_e(\theta_t) \in \mathcal{X}^i$, $u^i_e(\theta_t) \in \mathcal{U}^i$ for all $i = 1, \ldots, N$ such that $x^i_e = A^i x^i_e + B^i u^i_e$ and $C^i x^i_e = \theta_t$.

(A2)  The state and input constraint sets $\mathcal{X}^i$ and $\mathcal{U}^i$ contain all $x^i_e$ and $u^i_e$ equilibrium pairs in their interior, respectively, for all $i = 1, \ldots, N$.

Then, the closed-loop system (22) asymptotically converges to the set of equilibria $\mathcal{E}$ with domain of attraction $\mathcal{X}_0$.

*Proof:* We introduce the following notation:

$$J^i\left(x^i_t, U^i_t, \theta_t\right) = \sum_{k=0}^{T-1} V^i\left(x^i_{k,t}, u^i_{k,t}, \theta_t\right) \tag{23}$$

and

$$J(x_t, U_t, \theta_t) = \sum_{i=1}^{N} J^i\left(x^i_t, U^i_t, \theta_t\right). \tag{24}$$

The optimal value function obtained from solving problem (7) at time $t$ will thus be denoted as $J^*(x_t, U^*_t, \theta^*_t)$.

We will show first that the optimal value function $J^*(x_t, U^*_t, \theta^*_t)$ decreases along the closed-loop trajectories of the overall system at each time step $J^*(x_{t+1}, U^*_{t+1}, \theta^*_{t+1}) \leq J^*(x_t, U^*_t, \theta^*_t)$, if the assumptions of the theorem hold.

Let the initial state at time $t$ be $x_t = x_{0,t} \in \mathcal{X}_0$ and let $U^*_t = [u^*_{0,t}, \ldots, u^*_{T-1,t}]$ and $\theta^*_t$ be the optimizers of problem (7). Denote with $\mathbf{x}^*_t = [x_{0,t}, x^*_{1,t}, \ldots, x^*_{T,t}]$ the corresponding optimal state trajectory, with $1_N \otimes \theta^*_t = Cx^*_{T,t}$. Let $x_{t+1} = x^*_{1,t} = Ax_{0,t} + Bu^*_{0,t}$ and consider problem (7) for time $t + 1$. We will construct an upper bound for $J^*(x_{t+1}, U^*_{t+1}, \theta^*_{t+1})$. Consider the sequence $U_{t+1} = [u^*_{1,t}, \ldots, u^*_{T-1,t}, v]$ and the corresponding state trajectory resulting from the initial state $x_{t+1}$, $\mathbf{x}_{t+1} = [x^*_{1,t}, \ldots, x^*_{T,t}, Ax^*_{T,t} + Bv]$. The input $U_{t+1}$ will be feasible for the problem at $t + 1$ if and only if $v \in \mathcal{U}$ keeps $C(Ax^*_{T,t} + Bv)$ equal to some $1_N \otimes \theta$ with $\theta \in \Theta$ at step $T$ of the prediction, i.e., $C(Ax^*_{T,t} + Bv) = 1_N \otimes \theta$. Such $v$ exists by hypothesis (A1). Since $x^*_{T,t}$ is an equilibrium of the system, this also allows us to choose a feasible $v$, for which in fact $C(Ax^*_{T,t} + Bv) = 1_N \otimes \theta^*_t$. This is accomplished by noticing that $x^*_{T,t} = x_e(\theta^*_t)$ and selecting

$$v = u_e(\theta^*_t). \tag{25}$$

$J(x_{t+1}, U_{t+1}, \theta^*_t)$ will be an upper bound for the optimal $J^*(x_{t+1}, U^*_{t+1}, \theta^*_{t+1})$. Since trajectories generated by $U^*_t$ and $U_{t+1}$ overlap (except for the first and last sampling intervals), it is immediate to show that

$$
\begin{aligned}
&J^*(x_{t+1}, U^*_{t+1}, \theta^*_{t+1}) \\
\leq &J(x_{t+1}, U_{t+1}, \theta^*_t) \\
= &J^*(x_t, U^*_t, \theta^*_t) - (x_{0,t} - x_e(\theta^*_t))^\intercal Q(x_{0,t} - x_e(\theta^*_t)) \\
&- (u^*_{0,t} - u_e(\theta^*_t))^\intercal R(u^*_{0,t} - u_e(\theta^*_t)) \\
&+ ((Ax^*_{T,t} + Bv) - x_e(\theta^*_t))^\intercal Q((Ax^*_{T,t} + Bv) - x_e(\theta^*_t)) \\
&+ (v - u_e(\theta^*_t))^\intercal R(v - u_e(\theta^*_t)),
\end{aligned}
\tag{26}
$$

where $Q = \mathbf{diag}(Q^i) \in \mathbb{R}^{\sum_i n^i \times \sum_i n^i}$, $R = \mathbf{diag}(R^i) \in \mathbb{R}^{\sum_i m^i \times \sum_i m^i}$. Choosing the particular $v$ value given in (25) leads to $Ax^*_{T,t} + Bv - x_e(\theta^*_t) = 0$, so equation (26) becomes

$$
\begin{aligned}
&J^*(x_{t+1}, U^*_{t+1}, \theta^*_{t+1}) - J^*(x_t, U^*_t, \theta^*_t) \\
\leq &- (x_{0,t} - x_e(\theta^*_t))^\intercal Q(x_{0,t} - x_e(\theta^*_t)) \\
&- (u^*_{0,t} - u_e(\theta^*_t))^\intercal R(u^*_{0,t} - u_e(\theta^*_t)) \\
\leq &- \gamma(\|(x_t - x_e(\theta), u_t - u_e(\theta))\|), \quad \forall x_t \in \mathcal{X}_t.
\end{aligned}
\tag{27}
$$

where $\gamma$ is a class $K$ function. This inequality along with hypothesis (A0) on the matrices $Q$ and $R$ ensure that $J^*(x_t, U^*_t, \theta^*_t)$ decreases along the state trajectories of the closed-loop system (22) for any $x_t \in \mathcal{X}_t$. Since $J^*(x_t, U^*_t, \theta^*_t) \geq 0$ for all $x_t, U^*_t, \theta^*_t$, it follows that $J^*(x_t, U^*_t, \theta^*_t) \to J^\star$ as $t \to \infty$, where $J^\star$ is a nonnegative constant. We conclude that $J^*(x_{t+1}, U^*_{t+1}, \theta^*_{t+1}) -$ $J^*(x_t, U_t^*, \theta_t^*) \to 0$ as $t \to \infty$ and this implies that $\gamma(\|(x_t - x_e(\theta), u_t - u_e(\theta))\|) \to 0$. From $\gamma(\cdot)$ being a $K$ function, it follows that $x_t - x_e(\theta), u_t - u_e(\theta) \to 0$ as $t \to \infty$. $\square$

### B. Interrupted Negotiations

In case the distributed negotiation process is interrupted (e.g., due to execution time constraints) or otherwise allowed to run only for a finite number of iterations before the control inputs are implemented, the $\theta_t^i$ values do not converge to a common optimal value $\theta_t^*$. This means that individual agents will issue control commands that will guide them to possibly close but different terminal consensus points. In such a situation, we desire to find conditions under which repeated negotiation and implementation of intermediate consensus results will still allow asymptotic convergence to a *common* consensus point for each agent.

We propose an algorithm that fulfills the above objective if the subgradient iterations in subsequent time steps approach the optimal consensus point to an increasingly more accurate level and at the same time the local MPC solutions satisfy an improvement property along the closed-loop evolution of the agents' dynamics. The first requirement ensures that the mismatch between different interrupted $\theta_t^i$ values diminishes as $t \to \infty$. The second requirement is analogous to the standard suboptimal MPC scheme in [6], where it is established that feasibility of such an improvement constraint implies stability of the receding horizon control scheme.

In the following, we will denote the last (i.e., implemented) final consensus point reached by agent $i$ in the subgradient negotiation process of time instance $t$ by $\theta_t^i$. This intermediate consensus point is not optimal for the global optimization problem (7), but due to Assumptions 2-3 it is certainly feasible for the following local problem:

$$\min_{\theta_t^i} \quad q^i\left(x_t^i, \theta_t^i\right) \tag{28a}$$

$$\text{subj. to} \quad \theta_t^i \in \Theta. \tag{28b}$$

Distinguishing between the local $\theta_t^i$ variables allows the original global optimization problem (9) to be restated as

$$\min_{\theta_t} \quad \sum_{i=1}^{N} q^i(x_t^i, \theta_t^i) \tag{29}$$

$$\text{subj. to} \quad \theta_t^1 = \cdots = \theta_t^N \in \Theta.$$

As opposed to the fully converged subgradient scheme in Section IV-A, the $\theta_t^i$ variables do not converge to the globally optimal one, thus we cannot rely on optimality of the MPC scheme to prove global convergence. Instead, an improvement property as shown in [6], which is required for asymptotic convergence to the set of equilibria will be formulated as

$$\sum_{i=1}^{N} \left( J^i \left( x_{t+1}^i, U_{t+1}^i, \theta_{t+1}^i \right) - J^i \left( x_t^i, U_t^i, \theta_t^i \right) \right)$$

$$\leq -\gamma(\|(x_t - x_e(\theta), u_t - u_e(\theta))\|), \tag{30}$$

where $\gamma$ is a class $K$ function. A feasible sequence for such a constraint always exists based on Assumptions 2-3 and the earlier developments in Section IV-A.

The value function improvement property in (30) is not sufficient for convergence to the global optimum, since the *common* terminal point constraint is missing and the local $\theta_t^{i*}$ values are in general different. Thus, if the agents' initial states are close to their local $x_e^i(\theta_t^{i*})$ equilibria, which are significantly different from each other, then any subgradient-based or other adjustment of the local terminal point constraints towards the globally optimal $\theta_t^*$ value would necessarily result in both local and global cost increase.

This suggests that an additional requirement besides the cost improvement property is needed, which ensures that the $\theta_t^i$ values will also converge over time to a common $\theta_t$. This can be accomplished by requiring that in each iteration the subgradient-based negotiation scheme is executed at least until

$$\|\theta_t^i - \theta_t^*\| \leq \varepsilon_t \quad \forall i = 1, \dots, N, \tag{31}$$

where the approximation bound is updated for instance according to

$$\varepsilon_t \leq \epsilon \frac{1}{t}, \quad \epsilon > 0. \tag{32}$$

In order to have some information about the required number of incremental subgradient iterations that guarantee fulfillment of constraints (30) and (31), we will make use of the following result from [7]. It can be shown that under a strong convexity type assumption, the incremental subgradient method defined earlier in (12)-(14) with an appropriately chosen stepsize $\alpha(k)$ has a sublinear convergence rate:

*Proposition 1:* [7] Let Assumptions 4 and 5 hold, and assume that there exists a positive scalar $\mu$ such that

$$q(x_t, \theta_t) - q^*(x_t) \geq \mu \left( \mathbf{dist}(\theta_t, \Theta_t^*) \right)^2, \quad \forall \theta_t \in \Theta. \tag{33}$$

Then for the sequence $\{\theta_t(k)\}$ generated by the incremental subgradient method with the stepsize of $\alpha(k) = \frac{1}{2\mu}\frac{1}{k+1}$, $\mu > 0$, we have

$$(\mathbf{dist}(\theta_t(k+1), \Theta_t^*))^2 \leq \frac{1 + \ln(k+1)}{k+1}\frac{N^2\beta^2}{4\mu^2}. \tag{34}$$

In the following, we describe a scheme, which allows the two conditions (30) and (31) to be tested based on the cyclic communication scheme underlying the subgradient-based negotiation.

In Algorithm 1, agents perform cyclic iterations of the subgradient (SG) method (12)-(14). They execute at least the number of iterations dictated by the optimal $\theta_t^*$ approximation requirement in (31). Satisfaction of the test (31) is signaled by a flag $f_{\mathrm{SG}}$. If needed, agents continue with subgradient iterations until the global cost improvement property in (30) is satisfied. This is signaled by flag $f_{\mathrm{DMPC}}$.

In order to accomplish this, agents pass along besides their current subiterate $\vartheta_t^i(k)$ of the consensus point in iteration $k$ at sampling time $t$, the two binary variables (flags) $f_{\mathrm{DMPC}}$ and $f_{\mathrm{SG}}$ corresponding to tests (30) and (31), and two vectors of dimension $N$: $J_{\mathrm{curr}}$ and $J_{\mathrm{prev}}$. These vectors contain the individual cost values associated with the current and previous sampling time, respectively. $J_{\mathrm{prev}}$ has values corresponding to the cost of using the final $\vartheta_{t-1}^i(k)$ consensus points for implementation during the previous sampling time $t-1$. The current cost $J_{\mathrm{curr}}$ gets filled up cyclically using the most recent subiterate $\vartheta_t^i(k)$ for each agent.

When an agent computes its own consensus point subiterate, it calculates the corresponding local cost value and checks the sum of previous and current cost values for each agent to decide whether the improvement property (30) is satisfied. If it is, then it sets a flag $f_{\mathrm{DMPC}}$, which indicates that the improvement property (30) is fulfilled and every other agent should enter in an implementation phase, provided that condition (31) is also satisfied. The message reaches all other agents eventually as they pass along this information in a cyclic pattern. If property (30) is not satisfied, then it puts its current cost value entry in the vector $J_{\mathrm{curr}}$ and passes it on to the next agent.

*Theorem 2:* Under the assumptions of Section II, Algorithm 1 converges asymptotically to the set of equilibria $\mathcal{E}$.

*Proof:* The main idea of the proof follows along the lines of Theorem 1, except for two crucial points. A feasible sequence for the improvement constraint (30) always exists based on Assumptions 2-3 and the developments in Section IV-A. This improvement property guarantees that even with interrupted negotiations, the distributed MPC problem converges asymptotically to some set of different terminal points (since $\theta_t^i$ are different in this case). However, these terminal points are guaranteed to form a single consensus point, attained asymptotically by the repeated application of the iterative subgradient method, due to (31) and the compactness of $\Theta$. $\square$

*Remark 3:* Although Algorithm 1 guarantees global convergence, it requires an increasing number of subgradient iterations in subsequent time steps in order to approach the optimal value with a decreasing tolerance. The requirements (30) and (31) are only sufficient conditions and thus might be somewhat conservative. Decreasing the initial stepsize of the subgradient iterations may solve this problem. The increasing number of subgradient iterations can also be alleviated in practice in the following way: Once $\varepsilon_t$ gets small enough or another condition indicating closeness to the global consensus point is satisfied, the $\theta$ consensus point can be fixed for all agents and the scheme could proceed with a pure decentralized MPC scheme. This would ensure convergence due to the result shown in Section IV-A.

## V. Numerical Example of an Aerial Refueling Scenario

This numerical example considers a simplified aerial refueling scenario with three aircraft, which illustrates the importance of negotiating an optimal consensus point and the difference from standard rendezvous problems. The scheme involves the linearized longitudinal dynamics of three aircraft representing the tanker and two smaller aircraft to be serviced, respectively. The simplified objective is to control all three aircraft from their initial status to the same altitude and airspeed, which will allow the refueling process to be initiated. The consensus variable $\theta_t = [\Delta h \quad \Delta V]^\intercal$ will thus be a vector of dimension two, with entries of altitude and true airspeed deviations from the trim values, respectively. The optimal rendezvous altitude and airspeed will be determined using the distributed (cyclic) negotiation process based on local subgradient iterations. Once the corresponding control actions are implemented, the negotiations start again in the subsequent sampling time from the new initial states. This receding horizon procedure controls all three aircraft to a common altitude and airspeed. The optimal choice of the rendezvous parameters depend on the dynamics and constraints of the individual aircraft, along with the weighting matrices of the local cost functions, which penalize deviations in altitude, velocity and the use of control actions.

The tanker is represented by a Boeing 747-100/200 aircraft model based on [8]. The two serviced vehicles are modeled as F-16 aircraft from [9], [10]. The models were all discretized with a sampling time of $0.05$ s. The cost function of the tanker penalizes altitude-changes heavily, and the cost functions associated with the two fighters are based on their health or fuel level.

The linearized longitudinal models of all aircraft involved correspond to a straight and level flight condition at $4000$ m altitude and $184$ m/s true airspeed. The control inputs of the linearized Boeing 747-100/200 aircraft represent deviations from the trim values of thrust and elevator deflection, denoted by $u^{\mathrm{B747}} = [\delta_{th}^{\mathrm{B747}} \quad \delta_e]^{\mathsf{T}}$, respectively. The F-16 aircraft models are equipped with an inner-loop pitch rate controller so the control inputs of their linearized models are deviation from trim throttle and desired pitch rate command $u^{\mathrm{F16}} = [\delta_{th}^{\mathrm{F16}} \quad \delta_{qcmd}]^{\mathsf{T}}$, respectively. The constraints on the aircraft inputs are defined as the following

$$
\begin{bmatrix} -50000 \text{ lb} \\ -10 \text{ deg} \\ -1000 \text{ lb} \\ -100 \text{ deg/s} \end{bmatrix} \leq \begin{bmatrix} \delta_{th}^{\mathrm{B747}} \\ \delta_e \\ \delta_{th}^{\mathrm{F16}} \\ \delta_{qcmd} \end{bmatrix} \leq \begin{bmatrix} +150000 \text{ lb} \\ +10 \text{ deg} \\ +5000 \text{ lb} \\ +100 \text{ deg/s} \end{bmatrix}. \tag{35}
$$

The initial altitudes of the two F-16s are chosen as $\pm 30.48$ m around the trim altitude. The tanker's initial altitude w.r.t. the trim value is $-10$ m. All aircraft are initialized as flying straight and level with the same trim velocity.

If we simply considered the average value of the initial altitudes as a rendezvous point, then the aircraft should approach approximately $-3$ m. Due to the different dynamics and cost functions of the aircraft representing fuel cost and vehicle health priorities, this average value is not optimal. We consider the following weighting matrices in the local cost functions

$$
Q^{\mathrm{B747}} = \begin{bmatrix} 1 & 0 \\ 0 & 1 \end{bmatrix}, \quad R^{\mathrm{B747}} = \begin{bmatrix} 10^{-7} & 0 \\ 0 & 10^4 \end{bmatrix}, \tag{36}
$$

$$
Q_1^{\mathrm{F16}} = \begin{bmatrix} 10 & 0 \\ 0 & 10 \end{bmatrix}, \quad R_1^{\mathrm{F16}} = \begin{bmatrix} 10^{-5} & 0 \\ 0 & 0.5 \end{bmatrix}, \tag{37}
$$

$$
Q_2^{\mathrm{F16}} = \begin{bmatrix} 0.4 & 0 \\ 0 & 10 \end{bmatrix}, \quad R_2^{\mathrm{F16}} = \begin{bmatrix} 10^{-5} & 0 \\ 0 & 0.1 \end{bmatrix}. \tag{38}
$$

These weights intend to mimic a situation, where the F-16 indexed as #1 penalizes altitude changes and pitch control more heavily due to a restriction posed by limited fuel supply or elevator control authority.

Simulation results of the centralized solution to the optimal finite-time rendezvous problem with time horizon of $T = 100$ steps (5 seconds) are shown in Figure 1. The optimal rendezvous point is determined to be at 21.2340 m away from the trim altitude and the common terminal velocity at straight and level flight equilibrium is $-1.1933$ m/s away from the initial trim velocity. Notice that the rendezvous altitude is quite close to F-16 #1 as an indication of its limited maneuvering capabilities represented in its local cost function.

The problem can be solved in a distributed fashion using the incremental subgradient method described in [2] and implemented through a cyclic sequential update. Since the optimal rendezvous point and the corresponding control solutions converge to the centralized one asymptotically, only convergence of $\theta_t^i(k)$ are shown in Figure 3.

In order to have a receding horizon implementation of the incremental subgradient method proposed in Algorithm 1, the maximum number of major subgradient iterations were limited to 15 before applying the local control solutions at each new MPC update. The limited number of subgradient iterations lead to a mismatch between local versions of the common rendezvous point $\theta_t^i$. Still, the MPC implementation of the finite-time optimal rendezvous problem with local rendezvous point mismatch stabilizes the system and provides acceptable performance besides providing feedback action. The resulting trajectories are shown in Figure 2 until sampling time $t = 100$. The convergence of $\theta_t^{\text{B747}}(k)$ in each MPC update period is shown in Figure 4, where one horizontal axis represents the subgradient iterations $k$ (limited at 15) and the other horizontal axis corresponds to the MPC sampling time index $t$. Notice that besides the improvement made in each MPC sampling period, there is also a trend indicating that the local $\theta_t^{\text{B747}}(k)$ value starts from a closer point to $\theta_t^*(k)$ and approaches it more and more closely in subsequent MPC updates.

## VI. Conclusions

We have introduced a distributed model predictive control (DMPC) framework, where the control objective is to agree upon and achieve an optimal consensus point for constrained dynamic agents. The negotiation scheme makes use of the cyclic incremental subgradient algorithm described in [2]. Convergence properties of the combined DMPC / incremental subgradient approach were analyzed and a sufficient minimum number of subgradient iterations were established. An algorithm was proposed that ensures convergence of the decentralized scheme.

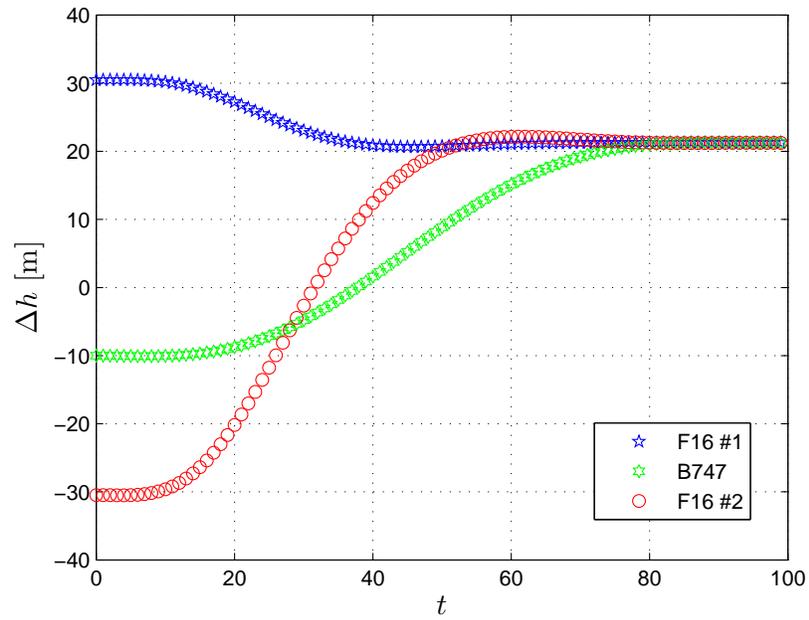

Fig. 1. Centralized solution of the finite-time optimal rendezvous problem in an aerial refueling scenario.

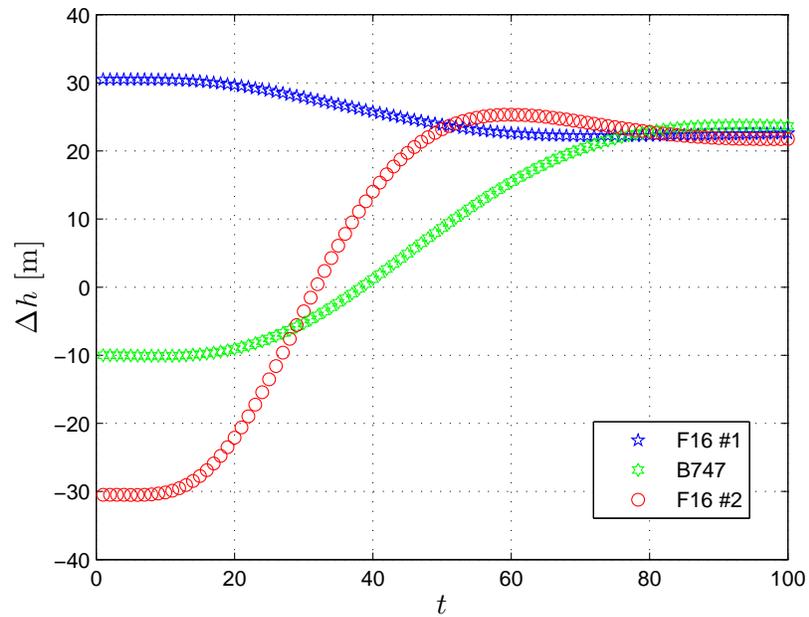

Fig. 2. MPC implementation of the finite-time optimal rendezvous problem with subgradient iterations limited at 15.

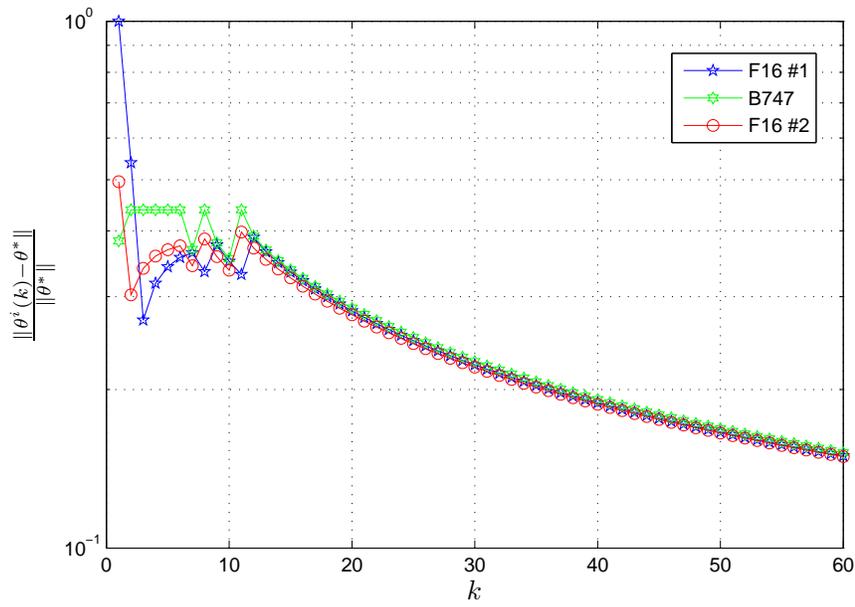

Fig. 3. Convergence of local estimates to the optimal rendezvous point using the incremental subgradient algorithm.

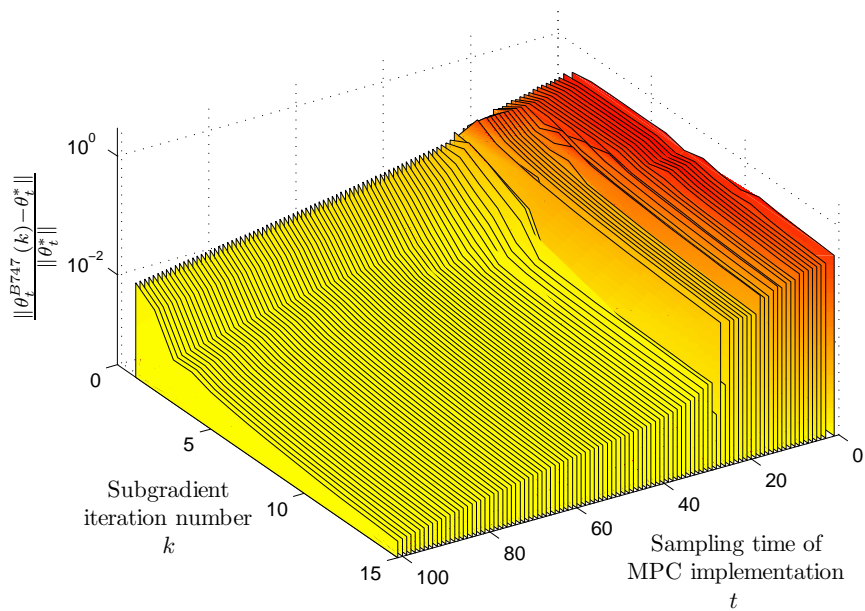

Fig. 4. Convergence of local estimates to the optimal rendezvous point during each MPC update.

Besides the aerial refueling numerical example shown in Section V, the approach proposed in this paper can be used in other distributed "synchronization" problems as well, where agents with constrained dynamics have to agree upon and achieve simultaneously an "optimal" consensus value, which is not known a priori. Our current work considers schemes that relax the cyclic, sequential communication requirement and rely on parallel, localized iterations.

## VII. ACKNOWLEDGEMENTS

The authors wish to thank Richard Murray for helpful discussions. The work of Tamás Keviczky was partially funded by the Boeing Corporation while he was a postdoctoral scholar at the Control and Dynamical Systems, California Institute of Technology. The work by Karl H. Johansson was partially funded by the Swedish Research Council, the Swedish Strategic Research Foundation, and the European Commission through the NoE HYCON. It was partially done during a stimulating sabbatical visit at Caltech, which is gratefully acknowledged.

---

**Algorithm 1**: Cyclic incremental DMPC algorithm

---

**1** Initialize $\beta, \mu, \epsilon, \theta_0^i$;

**2** $f_{\text{DMPC}}, f_{\text{SG}} \longleftarrow$ false;

**3** $k, t \longleftarrow 0$;

**4** $J_{\text{curr}} \in \mathbb{R}^N \longleftarrow \mathbf{0}$;

**5** $J_{\text{prev}} \in \mathbb{R}^N \longleftarrow -M \cdot \mathbf{1}$;                    /* M is large number */

**6** **loop**

**7**   Measure states $x_t^i$;

**8**   **repeat**

**9**     $\alpha(k) \longleftarrow \frac{1}{2\mu}\frac{1}{k+1}$;

**10**     $\vartheta_t^0(k) \longleftarrow \theta_t(k)$;

**11**     **for** $i = 1$ **to** $N$ **do**

**12**       **if** $f_{\text{DMPC}} \wedge f_{\text{SG}}$ **then**

**13**         $J_{\text{prev}}^i \longleftarrow J_t^i(x_t^i, U_t^i, \vartheta_t^i(k-1))$;

**14**         Implement $u_{0,t}^{i*}(\vartheta_t^i(k-1))$;

**15**         $\vartheta_t^i(k) \longleftarrow \vartheta_t^i(k-1)$

**16**       **else**

**17**         Compute a $g^i(k) \in \partial q^i(x_t^i, \vartheta_t^{i-1}(k))$;

**18**         $\vartheta_t^i(k) \longleftarrow \mathcal{P}_\Theta\left[\vartheta_t^{i-1}(k) - \alpha(k)g^i(k)\right]$;

**19**         $J_{\text{curr}}^i \longleftarrow J_t^i(x_t^i, U_t^i, \vartheta_t^i(k))$;

**20**         **if** $\sum_{i=1}^N (J_{\text{curr}}^i - J_{\text{prev}}^i) \leq 0$ **then**

**21**           Set $f_{\text{DMPC}}$ true;

**22**         **else**

**23**           Set $f_{\text{DMPC}}$ false;

**24**         **end**

**25**       **end**

**26**     **end**

**27**     $\theta_t(k+1) \longleftarrow \vartheta_t^N(k)$;

**28**     **if** $\frac{1+\ln(k+1)}{k+1}\frac{N^2\beta^2}{4\mu^2} \leq \frac{\epsilon}{t}$ **then**

**29**       Set $f_{\text{SG}}$ true;

**30**     **else**

**31**       Set $f_{\text{SG}}$ false;

**32**     **end**

**33**     $k \longleftarrow k+1$;

**34**   **until** *new measurement is available* ;

**35**   $t \longleftarrow t+1$;

**36**   $k \longleftarrow 0$;

**37** **end loop**

---